\documentclass[epj,twocolumn]{webofc}
\usepackage[varg]{txfonts}   

\begin{document}

\newcommand{\todo}[1]{\textbf{\textsc{\textcolor{red}{(TODO: #1)}}}}
\newcommand{\fcs}{Fe$_{1-x}$Co$_{x}$Si}
\newcommand{\mfs}{Mn$_{1-x}$Fe$_{x}$Si}
\newcommand{\mcs}{Mn$_{1-x}$Co$_{x}$Si}
\newcommand{\cso}{Cu$_{2}$OSeO$_{3}$}

\title{Neutron spin echo spectroscopy under 17\,T magnetic field at RESEDA}

\author{	J. Kindervater\inst{1}\fnsep\thanks{\email{jonas.kindervater@frm2.tum.de}} \and 
		N. Martin\inst{1,2} \and
		W. H\"au{\ss}ler\inst{1,2} \and
		M. Krautloher\inst{1,2}\and
		C. Fuchs\inst{2} \and
		S. M\"uhlbauer\inst{2} \and
		J. A. Lim\inst{2,3,4} \and\\
		E. Blackburn\inst{4} \and
		P. B\"oni\inst{1} \and
		C. Pfleiderer\inst{1}
		}

\institute{	Physik-Department, Technische Universit\"at M\"unchen, D-85748 Garching, Germany \and
		Heinz Maier-Leibnitz Zentrum (MLZ), Technische Universit\"at M\"unchen, D-85748 Garching, Germany\and
		Institut f\"ur Festk\"orperphysik, TU Dresden, D-01069 Dresden, Germany\and
		School of Physics and Astronomy, University of Birmingham, Birmingham B15 2TT, United Kingdom
          }

\abstract{
  We report proof-of-principle measurements at the neutron resonance spin echo spectrometer RESEDA (MLZ) under large magnetic fields by means of Modulation of IntEnsity with Zero Effort (MIEZE). Our study demonstrates the feasibility of applying strong magnetic fields up to 17\,T at the sample while maintaining unchanged sub-$\mu$eV resolution. We find that the MIEZE-spin-echo resolution curve remains essentially unchanged as a function of magnetic field up to the highest fields available, promising access to high fields without need for additional fine-tuning of the instrument. This sets the stage for the experimental investigation of subtle field dependent phenomena, such as magnetic field-driven phase transitions in hard and soft condensed matter physics. 
 }
\maketitle
\section{Introduction}
\label{intro}
A wide range of prominent scientific problems involving high magnetic fields, such as the spectrum of thermal fluctuations stabilising the Skyrmion lattice phase in chiral magnets \cite{muhlbauer2009skyrmion,georgii2011apl}, quantum phase transitions of transverse field Ising magnets \cite{Coldea} or field-induced Bose-Einstein condensation of magnons \cite{zheludev2007dynamics}, require neutron spectroscopy at sub-$\mu$eV resolution. Yet, despite this importance only very few studies of this kind have been reported in the literature. On the one hand, this situation may be traced to the limitations of conventional neutron scattering techniques such as triple axis spectroscopy (TAS) or time of flight spectroscopy (ToF), for which the resolution is directly tied to satisfying strict optical conditions causing a drastic loss of neutron intensity. On the other hand, it has long been established that the necessary high energy resolution may be achieved, in principle, by Neutron Spin Echo (NSE)\,\cite{mezei1972neutron}, which, however, requires non-depolarizing samples or sample environments. In turn, it has become an important instrumental challenge to extend neutron spin-echo spectroscopy towards depolarizing samples or sample environments.

As its key idea the NSE technique encodes the information on energy transfers in scattering events by the spin of the neutron. This permits complete decoupling of the energy resolution from the monochromaticity and -to some extent- the divergence of the neutron beam. In turn NSE reaches the highest energy resolution among all neutron spectroscopy techniques reported to date ($\delta E \sim$ 1\,neV), offering a dynamic range of typically 4 to 5 orders of magnitude. On the downside, being based on polarized neutrons it is crucial that the neutron polarization is not changed in an uncontrolled manner, completely prohibiting depolarizing conditions such as ferromagnetic or hydrogenated samples or the application of magnetic fields. 

A first approach to overcome the constraints of classical NSE has become known as Ferromagnetic NSE (FNSE) \cite{mezei1980principles} where adiabatic field transitions into the sample regions are used to conserve one polarization component while all others may be lost at the expense of at maximum one half of the polarization. However, due to its inherent complexity, this method, to the best of our knowledge, has only been used very rarely (see \emph{e.g.} \cite{boucher1985diffusion}). Further, if no component of the neutron beam polarization is conserved and the beam is depolarized completely at the sample, the spin echo signal will be destroyed. A second approach to overcome the limitations of NSE under such completely depolarizing conditions has become known as Intensity Modulation NSE \cite{farago1986study}. Representing a straight forward variant of FNSE, two additional spin polarizers before and after the sample are used to transform the phase modulation into an intensity modulation at the sample position. This intensity modulation is unaffected by the depolarization, getting transformed back into a phase modulation behind the sample. However, as its major drawback, this technique is subject to a large additional decrease in efficiency compared to NSE and FNSE. Namely, the two additional polarizers decrease the average intensity $I_{\mathsf{av}}$ by at least a factor of four and in the ferromagnetic echo the polarization $P$ is, by definition, reduced by another factor of 2. Hence, by comparison to NSE this technique is less efficient by at least a factor of $P^{2} \cdot I_{\mathsf{av}} \gtrsim 16$.

In contrast, the technique of Modulation of IntEnsity with Zero Effort (MIEZE)  \cite{gahler1992neutron}, which is based on the neutron resonance spin echo (NRSE) technique developed by G\"ahler and Golub \cite{gahler1987high,gahler1988neutron}, allows the combination of spin echo resolution with depolarizing conditions at high intensity. In this paper we report proof-of-principle measurements demonstrating the feasibility of using MIEZE as implemented at the NRSE spectrometer RESEDA (MLZ) for studies under high magnetic fields up to 17\,T at the sample position.

\section{The MIEZE technique}

\begin{figure}
\centering
\includegraphics[width=0.5\textwidth,clip]{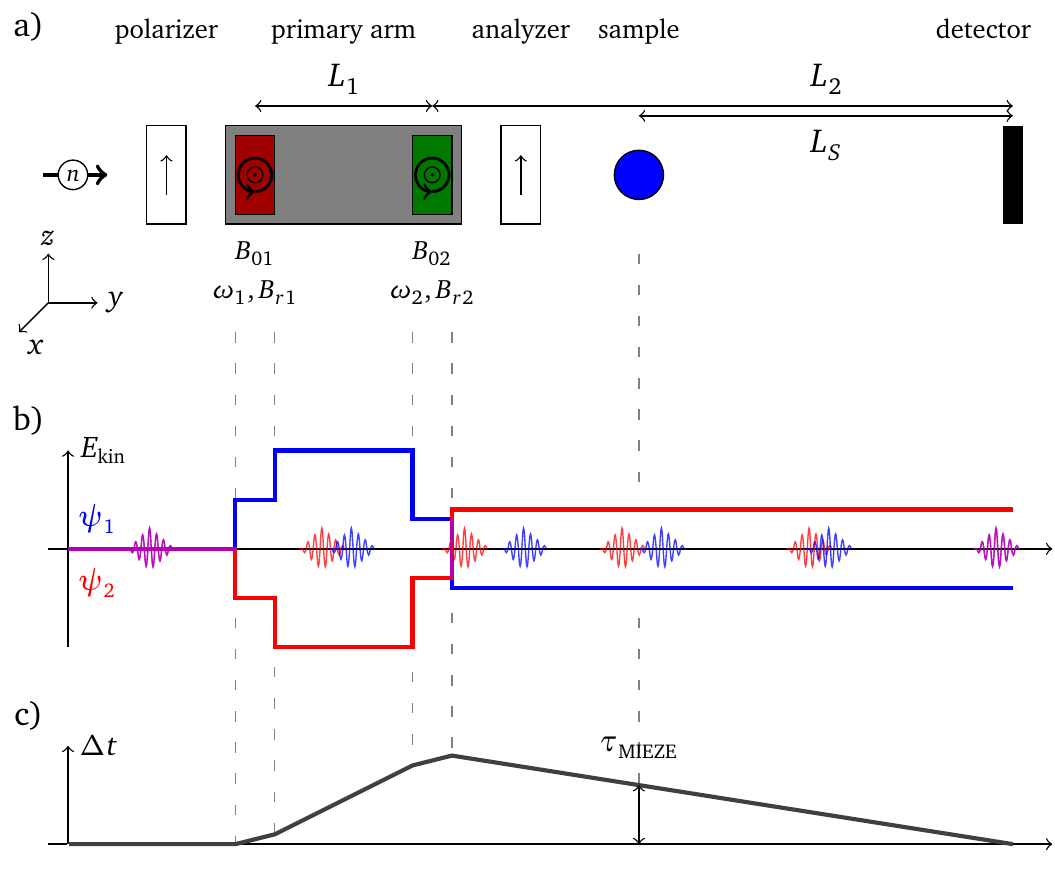}
\caption{Principles underlying the MIEZE technique. (a) Schematic depiction of a MIEZE-1 setup. (b) Kinetic energy of the neutron as function of flight path. (c) Temporal separation of the neutron spin wave function as function of the flight path. The MIEZE time $\tau_\textrm{MIEZE}$ is defined by the separation at the sample position.
}
\label{fig:MIEZE_I}       
\end{figure}

The MIEZE technique is conceptually based on NRSE, where resonant spin flippers comprised of a static magnetic $B_0$ and a perpendicular radio frequency field (RF-field) $B_r$ with frequency $\omega$ induce neutron spin precessions. For a pedagogical introduction we refer to \cite{MIEZE-reviews}. Shown in Fig.\,\ref{fig:MIEZE_I}\,(a) is a schematic depiction of a so-called MIEZE-1 setup consisting of a polarizer, two NRSE spin flippers, a spin analyzer and a time resolving detector. The first NRSE coil (cf red box in Fig.\,\ref{fig:MIEZE_I}\,(a)) introduces a splitting of the kinetic energy of the spin-up (red) and spin-down (blue) state of the neutron as shown in Fig.\,\ref{fig:MIEZE_I}\,(b). The spin states  accumulate a difference of time of flight, $\Delta t$, due to the difference in kinetic energy, when traveling the distance $L_1$ in the primary spectrometer arm. The second NRSE coil (cf. green box in Fig.\,\ref{fig:MIEZE_I}\,(a)) overcompensates the energy splitting such that this difference of time of flight returns to zero after the distance $L_2$ at the detector as depicted in Fig.\,\ref{fig:MIEZE_I}\,(c). The MIEZE time $\tau_\textrm{MIEZE}$ as the difference of time of flight at the sample position, being equivalent to the spin echo time in NSE (cf. Fig.\,\ref{fig:MIEZE_I}\,(a)), provides a measure of the resolution of the spectrometer. 

Apart form being insensitive to neutron depolarisation the MIEZE technique offers several additional advantages. First, being based on two NRSE coils as opposed to four coils in NRSE, the effort required for tuning when setting up the experiment is much reduced in MIEZE. Second, due to its lack of sensitivity on beam depolarisation and when no Bragg peak is accessible in the momentum range of interest the MIEZE technique permits to use strong incoherent scatterers (cf Vanadium or protonated samples) for measurements of the resolution function.

\begin{figure}
 \includegraphics[width=0.45\textwidth]{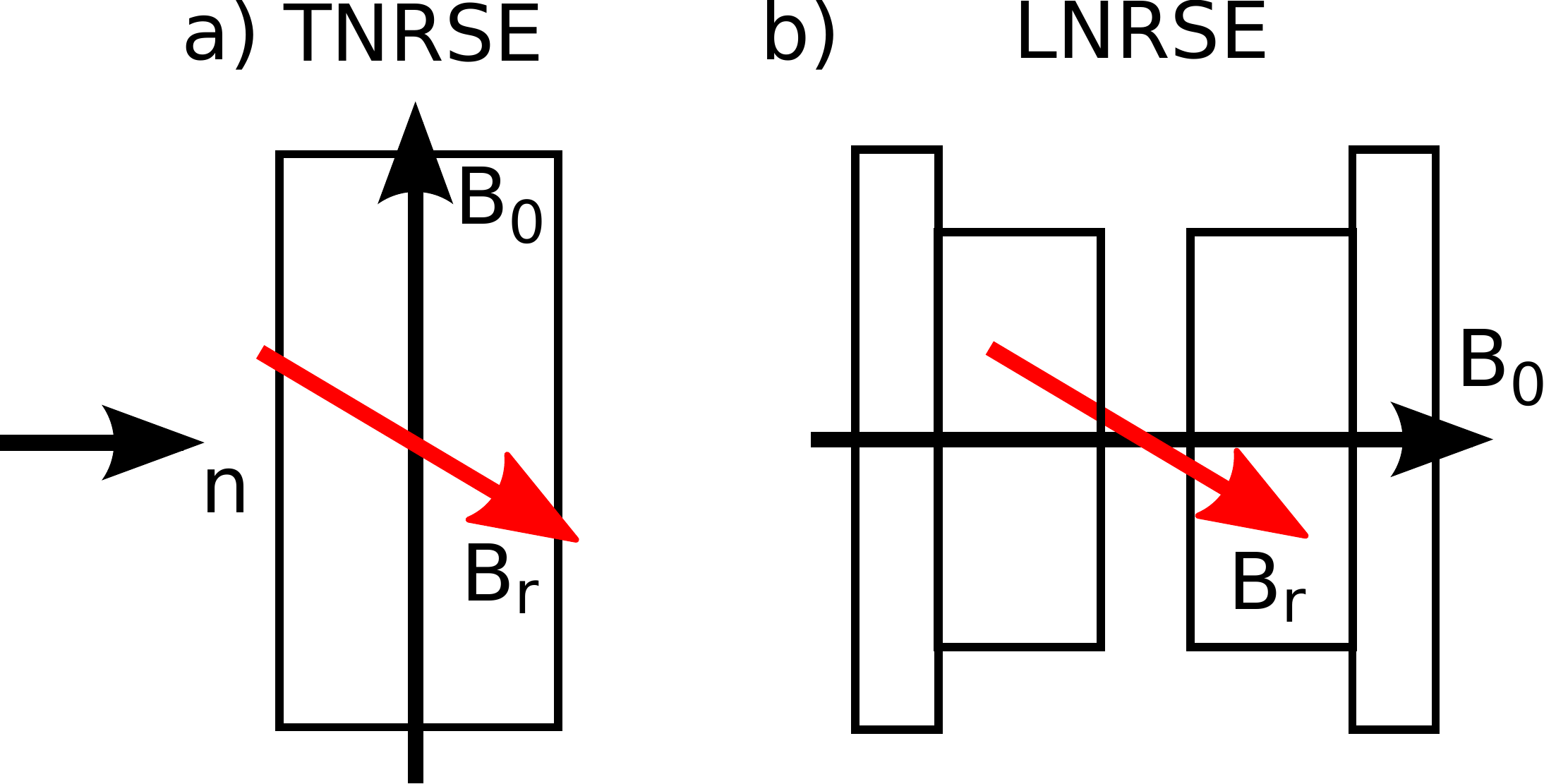}
 \caption{Schematic drawing of a TNRSE coil a) and LNRSE coil b). The neutron beam (n) traverses the coils from the left side as indicated by the arrow.}
 \label{fig:TNRSEvsLNRSE}
\end{figure}

\section{Implementation of MIEZE at RESEDA}

At the NRSE spectrometer RESEDA (MLZ) the MIEZE option has been implemented in terms of two different variants, a transverse and a longitudinal MIEZE setup referred to as t-MIEZE and l-MIEZE, respectively, which differ  in terms of the geometrical arrangement of $B_0$ and $B_r$ as shown in Fig.\,\ref{fig:TNRSEvsLNRSE}. In t-MIEZE  transverse NRSE (TNRSE) coils are used akin to the standard NRSE setup at RESEDA (c.f. Fig.\,\ref{fig:TNRSEvsLNRSE}\,(a)). In our experiments these coils were separated by $L_1=2.625$\,m. The $B_0$-fields in the TNRSE coils are \textit{perpendicular} to the neutron beam and the field region is defined accurately. 
The region between the TNRSE coils has to be field-free and is therefore shielded by a bespoke mu-metal housing. Coupling coils at the entrance and exit of the mu-metal shield control the transition of the polarization from the guide field of the polarizer into the zero field region and back into the guide field of the analyzer, where a compact V-cavity directly behind the mu-metal shielding is used as analyzer \cite{repper2012new}. The t-MIEZE setup covers already now a dynamic range of $0.1 \leq \tau_{\mathsf{MIEZE}} \leq 5$\,ns.

In contrast to t-MIEZE the l-MIEZE setup uses a field geometry (LNRSE) \cite{haussler2003neutron}, where two $B_0$-fields are \textit{parallel} to the neutron beam (c.f. Fig. \ref{fig:TNRSEvsLNRSE} b)) . In our experiments the distance was $L_1=1.925$\,m \cite{krautloher2014} with the same compact V-cavity as analyzer. The $B_0$-field is here generated by two solenoids in a Helmholtz geometry without accurately defined field boundaries. This has the significant advantage that magnetic shielding is not needed and longitudinal guide fields may be used to preserve the polarization. In addition, the similarity of the field geometry to classical spin echo allows to exploit the same correction techniques as in NSE, thus building on well-developed know-how. In its present form the LNRSE already extends the highest spin echo time accessible by a factor of 10 compared to TNRSE. Using effective field integral subtraction \cite{haussler2005effective} the lower limit for the spin echo time is less than 1\,ps and the dynamic range in this configuration covers four orders of magnitude.

The secondary spectrometer arm, which is the same for both setups, allows to perform measurements for sample-detector distances between 0.5 and 5\,m. A position sensitive CASCADE detector \cite{haussler2011detection,klein2011cascade} with an active area of 20\,cm $\times$ 20\,cm and a time resolution of $\Delta t=50$\,ns is used. In the CASCADE detector six boron foils (with thicknesses varying from 0.8 to 1.5\,$\mu$m) convert the neutrons, allowing to record a MIEZE signal with shifted phase on each foil and pixel. Combining the signals originating from all foils offers an efficiency of $\sim$ 90 \% of that of a standard $^{3}$He-based detector with a much larger detection depth.

\begin{figure}
 \includegraphics[width=0.45\textwidth]{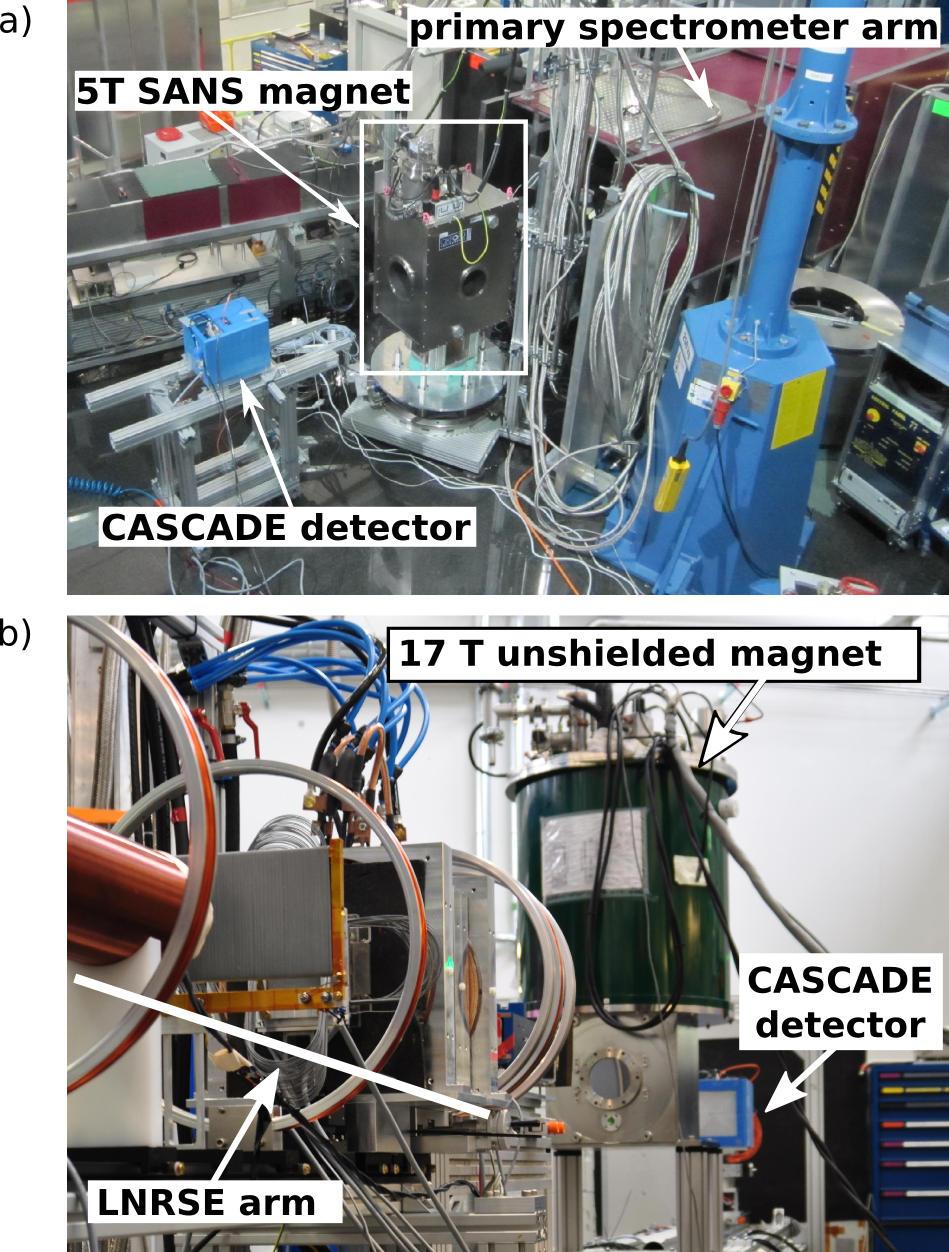}
 \caption{MIEZE and magnetic field setup at RESEDA. (a) Transverse MIEZE setup as combined with an actively shielded 5\,T SANS magnet by the MLZ sample environment group. (b) Longitudinal MIEZE setup as combined with the Birmingham 17\,T magnet \cite{holmes201217}. This magnet does not offer active stray-field reduction.}
 \label{fig:setup}
\end{figure}

\section{Experimental results}
For our proof of principle experiment two superconducting magnets have been used.  First a cryogen free,  5\,T magnet for small angle neutron scattering (SANS) with active stray-field compensation was used. This magnet may be set up for measurements with the magnetic field applied either longitudinal or perpendicular to the neutron beam. Second, a Helium cooled superconducting magnet for magnetic fields up to 17\,T  \cite{holmes201217} without stray field compensation. Setting up either magnet at RESEDA, including -slight- retuning the instrument requires approximately half a day.

\begin{figure}
\centering
\includegraphics[width=0.5\textwidth,clip]{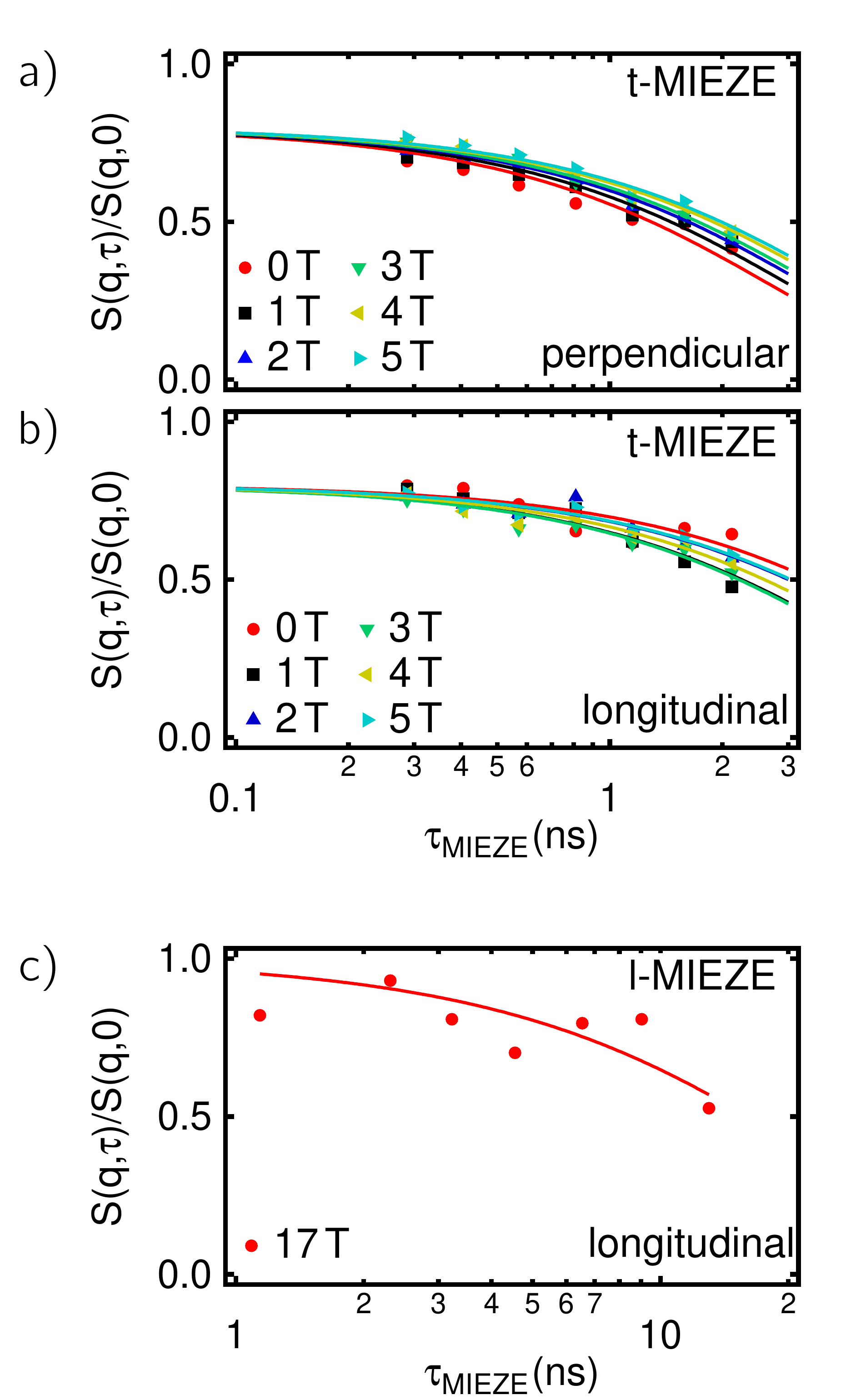}
\caption{MIEZE resolution with magnetic field at the sample. (a) Typical data as recorded with a 5\,T SANS magnet, where the field was applied perpendicular to the neutron beam. (b) same as for (a) for magnetic field applied longitudinal to the neutron beam. (c) Typical data as recorded at 17\,T applied longitudinal to the neutron beam using the Birmingham SANS magnet \cite{holmes201217}.}
\label{fig:Resolution}
\end{figure}

In a first series of tests the 5\,T SANS magnet was used with t-MIEZE option at a distance of $L_S=1.38$\,m to the detector and a distance between last TNRSE coil and detector of $L_2=3.08$\,m. The neutron wavelength was set to $\lambda=8.33$\,\AA. The results of a direct beam measurement are shown in figure \ref{fig:Resolution}\,(a) and (b). For the application of the field parallel to the neutron beam as combined with the t-MIEZE a solenoid around the spin analyzer was used to compensate the remaining stray fields disturbing the coupling of the neutron spin at the end of the magnetic shielding. In both geometries no reduction of the resolution between zero field and the maximum field of 5\,T was observed.

\begin{figure}
\centering
\includegraphics[width=0.5\textwidth,clip]{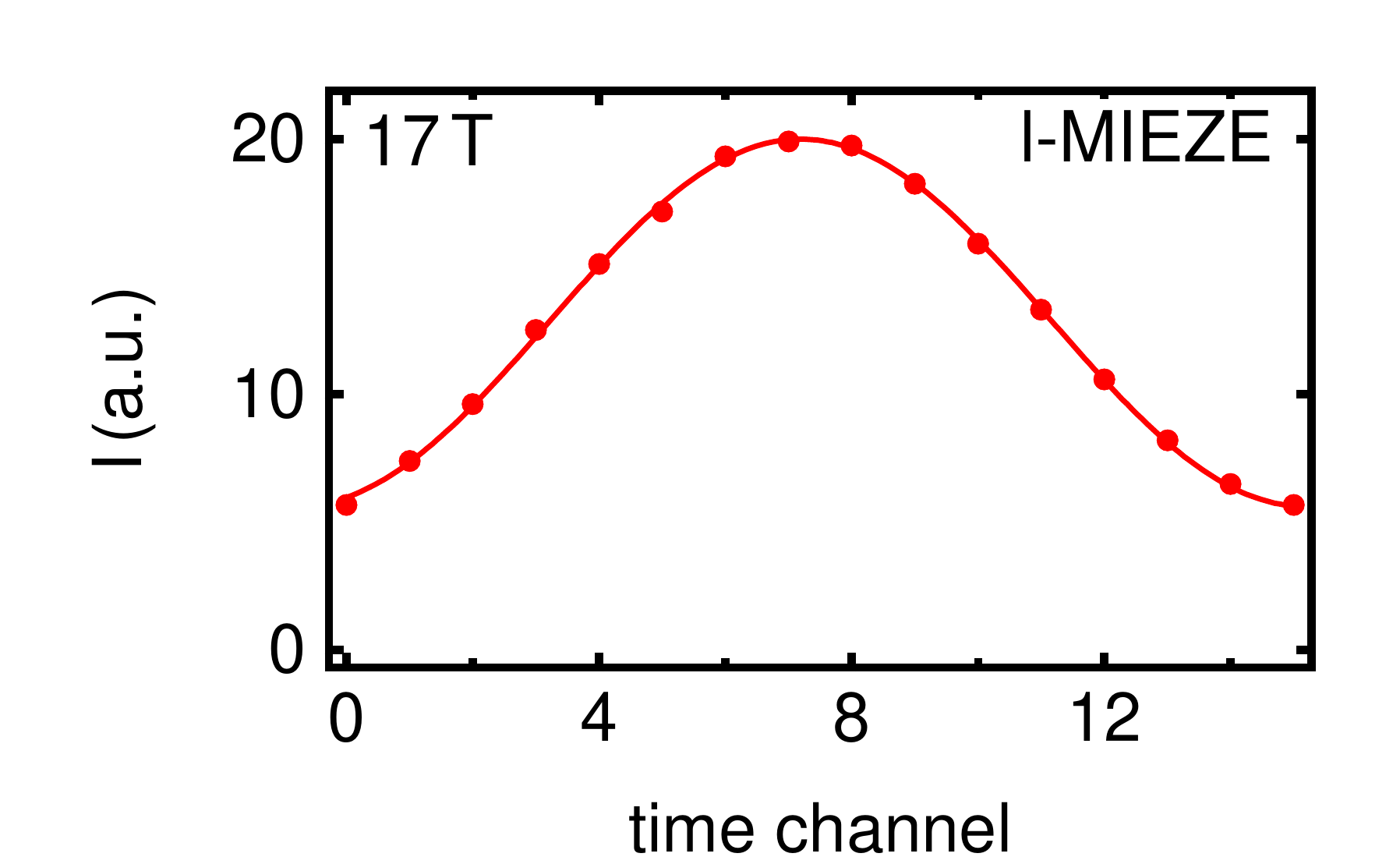}
\caption{Typical MIEZE echo measured with the l-MIEZE setup. The echo was recorded under an applied field of 17\,T at a frequency of $\omega=77$\,kHz. The line is cosine fit to the data.}
\label{fig:MIEZEEcho}
\end{figure}

The measurements with the 17\,T magnet were performed using the l-MIEZE setup as shown in figure \ref{fig:setup} b). As a general advantage of this setup, the l-MIEZE is less sensitive to the stray field of the magnet as described in further detail below. For our test neutrons with a wavelength of $\lambda=10.5$\,\AA\ were used. The magnet was placed at a distance of $L_S=2.6$\,m from the detector and with a distance of $L_2=5.00$\,m between the last LNRSE coil and the detector. Typical data from direct beam measurements, with 17\,T applied to the sample, are shown in Fig.\,\ref{fig:Resolution}\,(c). 

Shown in Fig.\,\ref{fig:MIEZEEcho} is a MIEZE echo with 60\% contrast as measured at 17\,T applied at the location of a sample. In this configuration, the large stray field of the non-compensated longitudinal magnetic field is compensated by a slight modification of the magnitude of the $B_0$ field produced in the second LNRSE coil. As seen in figure \ref{fig:Resolution} c), the achieved signal contrast at a Fourier time $\tau_{\mathsf{MIEZE}} \sim 15$\,ns is well above the commonly defined resolution limit of $1/e$. An extrapolation towards this limit suggests a resolution for this setup of about 20\,ns. In order to reach these high MIEZE times frequencies slightly above 1\,MHz and effective magnetic fields of about 70\,mT were applied in the LNRSE coils. 

\section{Conclusions and Outlook}
In conclusion we have demonstrated that large magnetic fields up to 17\,T may readily be combined with the MIEZE technique as implemented at RESEDA at MLZ. As the observed MIEZE resolution is independent from external conditions, field dependent studies are possible without the typical need for fine tuning of the instrument. We expect that these results promise access to a wide range of scientific questions in hard and soft condensed matter by means of high-resolution neutron spectroscopy. Last but not least the possibility to combine high-resolution neutron spectroscopy under high magnetic fields is also of great interest in the form of MIEZE as an add-on option for large scale SANS machines, bridging  characteristic times of quasi-elastic measurements and stroboscopic studies in the range $\Delta t \approx 1\,\mu\textrm{s}-1\,\textrm{ms}$ in addition, to SANS, TISANE and TAS/ToF.

\acknowledgement
We wish to thank the sample environment group of MLZ, notably J. Peters and H. Wei\ss, for assistance.
Financial support of DFG TRR80 and under ERC advanced grant 291079 are gratefully acknowledged.
JK acknowledges financial support by the TUM graduate school.
CP acknowledges funding by the BMBF under project {05K10WOC}.
EB acknowledges support by the EPSRC (EP/G027161/1 and EP/J016977/1).

\bibliography{BibDatenbank.bib}

\begin{thebibliography}{19}

\bibitem{muhlbauer2009skyrmion}
S.~M{\"u}hlbauer, B.~Binz, F.~Jonietz, C.~Pfleiderer, A.~Rosch, A.~Neubauer,
  R.~Georgii, P.~B{\"o}ni, Science \textbf{323}, 915 (2009)

\bibitem{georgii2011apl}
R.~Georgii, B.~Brandl, N.~Arend, W.~H\"au{\ss}ler, A.~Tischendorf,
  C.~Pfleiderer, P.~B\"oni, J.~Lal, Appl. Phys. Lett. \textbf{98}, 073505
  (2011)

\bibitem{Coldea}
R.~Coldea, D.A. Tennant, E.M. Wheeler, E.~Wawrzynska, D.~Prabhakaran,
  M.~Telling, K.~Habicht, P.~Smeibidl, K.~Kiefer, Science \textbf{327}, 177
  (2010)

\bibitem{zheludev2007dynamics}
A.~Zheludev, V.O. Garlea, T.~Masuda, H.~Manaka, L.P. Regnault, E.~Ressouche,
  B.~Grenier, J.H. Chung, Y.~Qiu, K.~Habicht et~al., Phys. Rev. B \textbf{76},
  054450 (2007)

\bibitem{mezei1972neutron}
F.~Mezei, Zeitschrift f{\"u}r Physik \textbf{255}, 146 (1972)

\bibitem{mezei1980principles}
F.~Mezei, \emph{The principles of neutron spin echo} (Springer, 1980)

\bibitem{boucher1985diffusion}
J.~Boucher, F.~Mezei, L.~Regnault, J.~Renard, Physical review letters
  \textbf{55}, 1778 (1985)

\bibitem{farago1986study}
B.~Farago, F.~Mezei, Physica B+ C \textbf{136}, 100 (1986)

\bibitem{gahler1992neutron}
R.~G{\"a}hler, R.~Golub, T.~Keller, Physica B: Condensed Matter \textbf{180},
  899 (1992)

\bibitem{gahler1987high}
R.~G{\"a}hler, R.~Golub, Zeitschrift f{\"u}r Physik B Condensed Matter
  \textbf{65}, 269 (1987)

\bibitem{gahler1988neutron}
R.~G{\"a}hler, R.~Golub, Journal de physique \textbf{49}, 1195 (1988)

\bibitem{MIEZE-reviews}
T.~Keller, R.~Golub, R.~G\"ahler, \emph{Neutron spin echo - a technique for
  high-resolution neutron scattering} (Academic Press, London, 2002), p. 1264

\bibitem{repper2012new}
J.~Repper, W.~H{\"a}u{\ss}ler, A.~Ostermann, L.~Kredler, A.~Chac{\'o}n,
  P.~B{\"o}ni, \textbf{340}, 012036 (2012)

\bibitem{haussler2003neutron}
W.~H{\"a}u{\ss}ler, U.~Schmidt, G.~Ehlers, F.~Mezei, Chemical physics
  \textbf{292}, 501 (2003)

\bibitem{krautloher2014}
M.~Krautloher, \textit{et al.} (in preperation)

\bibitem{haussler2005effective}
W.~H{\"a}ussler, U.~Schmidt, Physical Chemistry Chemical Physics \textbf{7},
  1245 (2005)

\bibitem{haussler2011detection}
W.~H{\"a}u{\ss}ler, P.~B{\"o}ni, M.~Klein, C.~Schmidt, U.~Schmidt, F.~Groitl,
  J.~Kindervater, Review of Scientific Instruments \textbf{82}, 045101 (2011)

\bibitem{klein2011cascade}
M.~Klein, C.J. Schmidt, Nuclear Instruments and Methods in Physics Research
  Section A: Accelerators, Spectrometers, Detectors and Associated Equipment
  \textbf{628}, 9 (2011)

\bibitem{holmes201217}
A.T. Holmes, G.R. Walsh, E.~Blackburn, E.M. Forgan, M.~Savey-Bennett, Review of
  Scientific Instruments \textbf{83}, 023904 (2012)

\end{thebibliography}
\end{document}